\documentclass{article}
\usepackage{amsfonts,amsmath,amssymb,mathrsfs}
\newcommand{\ket}[1]{\left| #1 \right>} % for Dirac bras
\newcommand{\bra}[1]{\left< #1 \right|} % for Dirac kets
\usepackage[latin1]{inputenc}
\usepackage[T1]{fontenc}
\usepackage[english]{babel}
\DeclareMathOperator{\e}{e}
\def\assign{\mathrel{:=}}
\def\diag{\mathop{\rm diag}}

\title{A qutrit Quantum Key Distribution protocol with better noise resistance}
\author{
  Fran\c{c}ois ARNAULT\\
  Universit\'e de Limoges --
  XLIM (UMR CNRS 7252)\\
  {\tt arnault@unilim.fr}
  \and
  Zo\'e AMBLARD\\
  Universit\'e de Limoges --
  XLIM (UMR CNRS 7252)\\
  {\tt zoe.amblard@unilim.fr}
}
\begin{document}
\maketitle

\bigskip

\bigskip

\bigskip

\begin{abstract}
The Ekert quantum key distribution protocol~\cite{Ekert1991} uses pairs of entangled qubits and
performs checks based on a Bell inequality to detect eavesdropping.
The 3DEB protocol~\cite{3DEB} uses instead pairs of entangled qutrits to achieve better noise
resistance than the Ekert protocol. It performs checks based on a Bell inequality for qutrits named
CHSH-3 and found in~\cite{CHSH3,ViolCHSH3GHZ}.  In this paper, we present a new protocol, which also uses pairs of entangled qutrits, but achieves even better noise resistance than 3DEB.  This gain of performance is obtained by using another inequality called here hCHSH-3, which was discovered in~\cite{Arnault2012}.  As the hCHSH3 inequality involve products of observables which become incompatible when using quantum states, we show how the parties running the protocol can measure
the violation of hCHSH3 in the presence of noise, to ensure the secrecy of the key. 
\end{abstract}
\newpage

\section{Introduction}

The Ekert91 protocol~\cite{Ekert1991} exploits pairs of entangled states to exchange keys, and uses Bell inequalities to detect eavesdropping.  Some of the measurement results obtained by the two parties Alice and Bob are perfectly correlated, providing key bits.  Other measurement results must exhibit quantum behavior if there is no alteration of the quantum channel, and this permits to detect evesdropping by testing a Bell inequality violation.

The amount of quantum violation is an important characteristic in key distribution protocols because larger violations leads to a better noise resistance~\cite{ViolAndNoise}.   Some progress has been made to increase this amount of violation with the use of parties with higher dimension (\cite{ViolStrongThan} and~\cite{Q3}, for qutrits).  The choice of the Bell inequality used to detect evesdropping is another parameter which can be considered. 

In their article introducing the 3DEB protocol~\cite{3DEB}, Durt, Cerf, Gisin and \^{Z}ukowski choosed to use three-dimensional quantum systems (qutrits), and the Bell inequality for qutrits named CHSH-3.   This way, they obtained better noise resistance than for the Ekert'91 protocol.

Our work makes one step further by using a recent discovered Bell inequality (here called hCHSH-3), which belongs to the family of homogeneous Bell inequalities introduced in~\cite{Arnault2012}.  The amount of violation which can be achieved with entangled states is even better than for CHSH-3. Consequently, the protocol we derive is more tolerant to noise, with a threshold of noise $F\simeq0.409$, instead of $F\simeq0.304$ for 3DEB.

Devices called multiport beam splitters~\cite{Tritters} (or tritters), are mentioned in~\cite{3DEB} as one way to handle measurements of qutrits.  Tritters are analyzed in~\cite{tritter1} and experimentally tested in~\cite{tritter2}.  Our new protocol h3DEB described in this article is analysed in view of the use of tritters to implement measurements.  A crucial point here will be that some products of observables, each implemented with tritters, can also be implemented by another single tritter.  This is needed for our protocol as the inequality hCHSH-3 involves such products. 

The paper is organized as follows.
It begins with some reminders and precisions about measurements with tritters in Section~2.  Then Section~3 recalls the 3DEB protocol and the CHSH-3 Bell inequality used by it.   After that, Section~4 introduces the Bell inequality hCHSH-3 we use, then considers the use of tritters for implementing the product of observables, and defines our new protocol h3DEB.  Finally, the paper concludes about the advantage of h3DEB providing better resistance to noise.
\newpage

\section{Prerequisites}

The 3DEB protocol and our protocol use qutrits and trichotomic observables.  For readability, we
assume that the outcomes of these observables are $1,\omega,\omega^2$ where $\omega$ is the third
root of unity ${\omega=e^\frac{2i\pi}{3}}$.  The observables used by the two parties Alice and Bob
will be denoted respectively by $A_i$ and~$B_j$ for some indexes $i$ and~$j$.  We will also use the
correlation functions introduced in~\cite{tritter2}~:
$$
  E(A_iB_j) = \sum_{a,b=1,\omega,\omega^2}  P(A_i=a,B_j=b) \, ab.
$$

\subsection{Measurements with tritters}

A tritter is parameterized by a triplet $(\varphi_0,\varphi_1,\varphi_2)$ of phase shifts.  For
readability we put $\theta_j=\exp(i\varphi_j)$ (for $j=0,1,2$) and
$\Theta=(\theta_0,\theta_1,\theta_2)$.  The tritter performs over a qutrit the following unitary
transformation : 
$$
  U_{\Theta} \assign HD_\Theta
  =
  \frac{1}{\sqrt3}\sum\limits_{k,l=0}^{2}\omega^{kl}\theta_{l}\ket{k}\bra{l}
$$
where the matrices $H$ and $D_\Theta$ are $H=(\omega^{kl})_{0\leq k,l\leq2}$ and
$D_\Theta=\diag(\theta_0,\theta_1,\theta_2)$.

In the specific and usual case where the phase shifts obey to the relation $\theta_{j}=\theta^{j}$, we have :
$$
 U_{\Theta}
 =\frac{1}{\sqrt3}\sum\limits_{k,l=0}^{2}\omega^{kl}\theta^{l}\ket{k}\bra{l}=\frac{1}{\sqrt3}\begin{pmatrix}
   1 & \theta & \theta^{2} \\
   1 & \omega\theta & \omega^{2}\theta^{2} \\
   1 & \omega^{2}\theta & \omega\theta^{2}
\end{pmatrix}
$$

  After the transformation performed by the tritter, a measurement is made using three detectors.
This measurement is represented by the observable
$$
  Z = \sum_{k=0}^{2} \omega^{k}\ket{k}\bra{k}.
$$
(Note that, as we assumed the three possible outcomes to be labeled by complex roots of unity, we
use unitary observables).  Thus, the measurement obtained by the combination of the tritter and
the detectors corresponds to the following observable 

\begin{equation}
\begin{split}
Z_{\Theta} \assign D_{\Theta^{*}}H^{\dagger}ZHD_{\Theta}=\begin{pmatrix}
   0 & 0 & \theta_{2}\theta_{0}^{*} \\
   \theta_{0}\theta_{1}^{*} & 0 & 0 \\
   0 & \theta_{1}\theta_{2}^{*} & 0
\end{pmatrix}.
\end{split}
\label{eq:ZTheta}
\end{equation}

which gives us, in the particular case where $\theta_{j}=\theta^{j}$:

\begin{equation}
\begin{split}
Z_{\Theta}=\begin{pmatrix}
   0 & 0 & \theta^{2} \\
   \theta^{*} & 0 & 0 \\
   0 & \theta^{*} & 0
\end{pmatrix}.
\end{split}
\label{ref1}
\end{equation}

\section{The 3DEB protocol}

We will recall the 3DEB protocol introduced in~\cite{3DEB}.  We begin with the CHSH-3 inequality
(or CHSH for qutrits) as defined in~\cite{CHSH3, ViolCHSH2Q3}, which is used for 3DEB.

\subsection{The inequality CHSH-3}

The CHSH-3 inequality can be written
\begin{equation*}
S \leqslant 2
\end{equation*}
where
\begin{small}
$S=\text{Re}\big(E(A_{1}B_{1})+E(A_{1}B_{2})-E(A_{2}B_{1})+E(A_{2}B_{2})\big)\\
\phantom{espacegrand}+\frac{1}{\sqrt{3}}\text{Im}\big(E(A_{1}B_{1})-E(A_{1}B_{2})-E(A_{2}B_{1})+E(A_{2}B_{2})\big).$
\end{small}
\bigskip

 Some entangled states are known to violate this inequality.  The GHZ state
\begin{equation}
\begin{split}
\ket{\psi}=\frac{1}{\sqrt3}(\ket{00}+\ket{11}+\ket{22})
\label{GHZstate}
\end{split}
\end{equation}
is known to violate CHSH-3 with a violation factor (the quotient of the quantum value with the classical bound) ${v=(6+4\sqrt3)/9\simeq 1.436}$.

 This violation factor is considered very important for the security of the key distribution protocol.  The presence of noise is usually modelized by the replacement of the initial entangled state by
a mixture
$$
  F \frac Id + (1-F) \ket\psi\bra\psi
$$
where $F$ is the proportion of noise.  The point is that the presence of noise decreases the experienced violation to $(1-F)v$ and that the protocol is considered useless when the initial state entanglement cannot be detected anymore.  With this criterion, it has been shown that the protocol 3DEB is resistant to the presence of noise up to a threshold $F=1-1/v=(11-6\sqrt3)/2\simeq 0.304$~\cite{Q3, UptoN16, ViolCHSH3GHZ}.

  The bases considered in most papers~\cite{ADGL, Q3, UptoN16} to obtain these best violations are the following four ``optimal bases'' (two for each party) corresponding to tritter measurements using the following phase shift triples $\Theta=(1,\theta,\theta^2)$ were $\theta$ is a suitable
power of $\zeta=\e^{\frac{2i\pi}{12}}$:
\expandafter\ifx\csname graph\endcsname\relax
   \csname newbox\expandafter\endcsname\csname graph\endcsname
\fi
\ifx\graphtemp\undefined
  \csname newdimen\endcsname\graphtemp
\fi
\expandafter\setbox\csname graph\endcsname
 =\vtop{\vskip 0pt\hbox{%
\pdfliteral{
q [] 0 d 1 J 1 j
0.576 w
0.072 w
q 0 g
28.8 -38.736 m
36 -40.536 l
28.8 -42.336 l
28.8 -38.736 l
B Q
0.576 w
0 -40.536 m
28.8 -40.536 l
S
Q
}%
    \graphtemp=.5ex
    \advance\graphtemp by 0.563in
    \rlap{\kern 0.650in\lower\graphtemp\hbox to 0pt{\hss $A_1$\hss}}%
\pdfliteral{
q [] 0 d 1 J 1 j
0.576 w
0.072 w
q 0 g
12.816 -14.688 m
18 -9.36 l
15.984 -16.488 l
12.816 -14.688 l
B Q
0.576 w
0 -40.536 m
14.4 -15.624 l
S
Q
}%
    \graphtemp=.5ex
    \advance\graphtemp by 0.000in
    \rlap{\kern 0.325in\lower\graphtemp\hbox to 0pt{\hss $A_2$\hss}}%
\pdfliteral{
q [] 0 d 1 J 1 j
0.576 w
0.072 w
q 0 g
24.048 -24.552 m
31.176 -22.536 l
25.848 -27.72 l
24.048 -24.552 l
B Q
0.576 w
0 -40.536 m
24.912 -26.136 l
S
Q
}%
    \graphtemp=.5ex
    \advance\graphtemp by 0.238in
    \rlap{\kern 0.563in\lower\graphtemp\hbox to 0pt{\hss $B_1$\hss}}%
\pdfliteral{
q [] 0 d 1 J 1 j
0.576 w
0.072 w
q 0 g
25.848 -53.352 m
31.176 -58.536 l
24.048 -56.52 l
25.848 -53.352 l
B Q
0.576 w
0 -40.536 m
24.912 -54.936 l
S
Q
}%
    \graphtemp=.5ex
    \advance\graphtemp by 0.888in
    \rlap{\kern 0.563in\lower\graphtemp\hbox to 0pt{\hss $B_2$\hss}}%
    \hbox{\vrule depth0.888in width0pt height 0pt}%
    \kern 0.650in
  }%
}%

\begin{figure}[h]
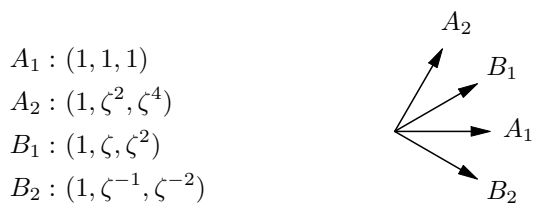

%\centering
\begin{minipage}[c]{0.5\linewidth}
\begin{align*}
  A_1: &\ (1, 1, 1) \\
  A_2: &\ (1, \zeta^2, \zeta^4) \\
  B_1: &\ (1, \zeta, \zeta^2) \\
  B_2: &\ (1, \zeta^{-1}, \zeta^{-2})
\end{align*}
\end{minipage}
\begin{minipage}[c]{0.5\linewidth}
\qquad\box\graph
\end{minipage}
\caption{\label{pic:hresult1} Representation of the four optimal bases in a plane parametrized by~$\theta$}
\end{figure}

  In~\cite{3DEB}, it is remarked that the phases used for $B_2$ can
be $(1,\zeta^3,\zeta^6)$ instead, as this change merely consists in reorder the three vectors of this base.

It was also remarked in~\cite{ADGL} that the non maximally entangled state
\begin{equation}
\frac{1}{\sqrt{2+\gamma^{2}}}(\ket{00}+\gamma\ket{11}+\ket{22})
\qquad\hbox{with $\gamma=\frac{\sqrt{11}-\sqrt{3}}{2}$}
\label{NMEstate}
\end{equation}
achieves an even better violation of CHSH-3, with a factor equal to ${\frac{1+\sqrt{\frac{11}{3}}}{2}\simeq 1.457}$.  This allows to reach a noise resistance up to the threshold $F\simeq 0.314$.

The expression of $S$ was rewritten in~\cite{Arnault2012} in the following form
\begin{equation}
S=-\frac{2}{9}\text{Re}(T)
\label{InegS}
\end{equation}
with
\begin{center}
$T=3\big((\omega^{2}-1)E(A_{1}^{2}B_{1}^{2}) + (\omega-1)E(A_{1}^{2}B_{2}^{2}) + (1-\omega^{2})E(A_{2}^{2}B_{1}^{2}) + (\omega^{2}-1)E(A_{2}^{2}B_{2}^{2})\big)$
\end{center}
Note that, the measurements outcomes being $1,\omega,\omega^2$, it is equivalent to consider the square of an observable and its complex conjugate.  For completeness, a derivation of this formula
is given in Appendix~A.   Now, the CHSH-3 inequality can be rewritten as
\begin{equation}
\text{Re}(-T) \leqslant 9.
\label{inegT}
\end{equation}
This last formulation will be useful to compare with the inequality we will use for our new
protocol.  Note that it was also shown in~\cite{Arnault2012} that the inequality obtained for qudits
($d\geq3$) in~\cite{ViolCHSH3GHZ} is the same as CHSH3 for the special case $d=3$.

\subsection{The 3DEB procedure}

Alice uses four observables $A_a$ with $a=0$ to 3, corresponding to tritter measurements
with phase shift triples $(1, \zeta^a, \zeta^{2a})$.  Bob use four observables $b_b$ with $b=0$ to
3, corresponding to tritter measurements with phase shift triples $(1,\zeta^{-b},\zeta^{-2b})$.
The following steps are repeated until Alice and Bob obtained a shared key of desired length.

\begin{enumerate}
\item Alice and Bob obtain an entangled pair of states in the GHZ state defined in~(\ref{GHZstate}).

\item Alice draws randomly a value for $a \in \{0,1,2,3\}$ and makes the measurement corresponding to the observable $A_{a}$ whereas Bob draws randomly a value for $b\in \{0,1,2,3\}$ and makes the measurement corresponding to the observable $B_{b}$. 

\item When $a=b$, the results obtained by Alice and Bob are perfectly correlated. Indeed, the two
tritters used by Alice and Bob perform on the shared GHZ state the transformation ${(H \otimes H)(D_{\Theta} \otimes D_{\Theta^{*}})}$, with $\Theta=(1,\zeta^a,\zeta^{2a})$.  But it is easy to check that:
\begin{equation}
{(H \otimes H)(D_{\Theta} \otimes D_{\Theta^{*}})(\ket{00}+\ket{11}+\ket{22})=(\ket{00}+\ket{12}+\ket{21})}
\label{CollectTrits}
\end{equation}
Consequently, in this case where $a=b$, Alice and Bob obtain a new trit for the shared key.

\item When $(a,b) \in \{(0,1), (0,3), (2,1), (2,3)\}$, Alice and Bob can use their joint measurements to detect eavesdropping, because the four observables $A_{0}, A_{2}, B_{1}, B_{2}$ correspond to a configuration of maximal violation of CHSH-3.  The same is true when $(a,b)\in\{(1,0),(1,2),(3,0),(3,2)\}$.
\end{enumerate}

These different cases can be summarized in Table~1.
\begin{table}[h]
\begin{minipage}[h]{0.5\linewidth}
\centering
\begin{tabular}{c | c c c c}
\hline\hline
    & $B_{0}$ & $B_{1}$ & $B_{2}$ & $B_{3}$\\
\hline
$A_{0}$ & $k$ & $c_{1}$ & $ $ & $c_{1}$\\ 
$A_{1}$ & $c_{2}$ & $k$ & $c_{2}$ & $ $\\ 
$A_{2}$ & $ $ & $c_{1}$ & $k$ & $c_{1}$\\ 
$A_{3}$ & $c_{2}$ & $ $ & $c_{2}$ & $k$\\ 
\end{tabular}
\label{tab:hresult1}
\caption{Cases for the 3DEB protocol}
\end{minipage}
\hfil
\begin{minipage}[h]{0.45\linewidth}
$k$ : Alice and Bob obtain key trits.\par
$c_{i}$ : Alice and Bob obtain values for two sets of data which can be used to check CHSH-3 violation.
\end{minipage}
\end{table}

When used with the maximally entangled GHZ state $\ket{\psi}$, the violation factor of CHSH-3 observed using this protocol (in the absence of noise) is equal to ${v=\frac{6+4\sqrt3}{2}\simeq1.436}$.  This corresponds to a noise resistance up to a threshold $F \simeq 0.304$.  Our aim was to create a new protocol more tolerant to noise, with a threshold $F$ greater than $0.304$.

\section{The new h3DEB protocol}
 
We will now describe our protocol.  It achieves better noise resistance because it will use an
homogeneous Bell inequality, which has a larger violation factor than CHSH-3.

\subsection{The homogeneous inequality}

  We will use the inequality :
\begin{equation}
 -2\text{Re}(T_{1}) \leqslant 9
\label{HBI}
\end{equation}
with
% This one for ADGL bases :
$
 T_{1}
  = (4\omega+2)E(A_1^2B_1^2) + (\omega-1)E(A_1^2B_1B_2) + (4\omega-1)E(A_1^2B_2^2) \\
\phantom{espace}-(2\omega+1)E(A_1A_2B_1^2) + (\omega-1)E(A_1A_2B_1B_2) + (\omega+2)E(A_1A_2B_2^2) \\
\phantom{espace}+(\omega+5)E(A_2^2B_1^2) + (\omega-1)E(A_2^2B_1B_2) - (2\omega+4)E(A_2^2B_2^2). \\
$
% The following is prefered if using 3DEB bases~:
%$T_{1}=-(2\omega+4)E(A_{1}^{2}B_{1}^{2}) + (\omega-1)E(A_{1}^{2}B_{1}B_{2}) + (4\omega+2)E(A_{1}^{2}B_{2}^{2}) \\
% \phantom{espace} + (\omega-1)E(A_{1}A_{2}B_{1}^{2}) - (2\omega+1)E(A_{1}A_{2}B_{1}B_{2}) + (4\omega-1)E(A_{1}A_{2}B_{2}^{2}) \\ 
% \phantom{espace} + (\omega+5)E(A_{2}^{2}B_{1}^{2})+ (\omega+2)E(A_{2}^{2}B_{1}B_{2}) + (\omega-1)E(A_{2}^{2}B_{2}^{2})$
%
\smallskip

This inequality belongs to the set of homogeneous Bell inequalities described in~\cite{Arnault2012}.
It has been shown in this paper that these inequalities are satisfied under the hypothesis of
local realism, and that they form a complete set.

A feature of the homogeneous Bell inequalities is that they involve some products (namely $A_1A_2$
and $B_1B_2$) of observables which become incompatible when considered as quantum observables.  The
outcomes of such a product of course cannot be meant to be the products of outcomes of incompatible
observables.  But if we use the unitary observables $Z_\Theta$ defined above for the $A_i$ and~$B_j$, the product of them is also a unitary observable which outcomes can be obtained with a single measurement.  Moreover, we argue here that this single measurement can be implemented by a slightly modified tritter.

\subsection{Products of incompatible observables}

%Previously, we have described tritters and detectors in the case where our triplets of phase shifts were $(1, \theta, \theta^{2})$, with $|\theta|=1$. To simulate our incompatible observables, we need to use triplet of phase shifts $(\gamma_{0}, \gamma_{1}, \gamma_{2})$, not necessarily powers of $\gamma$, but with $|\gamma_{i}|=1$ for $i=0,1,2$.

 Suppose that we have two tritters, which implement the observables $Z_\Theta$ and~$Z_\Lambda$
described by Equation~(\ref{eq:ZTheta}), with $\Theta=(\theta_0,\theta_1,\theta_2)$ and 
$\Lambda=(\lambda_0,\lambda_1,\lambda_2)$.  Then we need to implement the product observable
$Z_\Theta Z_\Lambda$.  But
\begin{small}
\begin{equation*}
\begin{split}
Z_\Theta Z_\Lambda=\begin{pmatrix}
   0 & 0 & \theta_{2}\theta_{0}^{*} \\
   \theta_{0}\theta_{1}^{*} & 0 & 0 \\
   0 & \theta_{1}\theta_{2}^{*} & 0
\end{pmatrix}
\begin{pmatrix}
   0 & 0 & \lambda_{2}\lambda_{0}^{*} \\
   \lambda_{0}\lambda_{1}^{*} & 0 & 0 \\
   0 & \lambda_{1}\lambda_{2}^{*} & 0
\end{pmatrix}
=
\begin{pmatrix}
   0 & \gamma_{0}^{*}\gamma_{1} & 0 \\
   0 & 0 & \gamma_{1}^{*}\gamma_{2} \\
   \gamma_{2}^{*}\gamma_{0} & 0 & 0
\end{pmatrix}
\end{split}
\end{equation*}
\end{small}
where
$$
  (\gamma_0,\gamma_1,\gamma_2)
  =
  (\theta_{2}^{*}\lambda_{1}^{*}, \theta_{0}^{*}\lambda_{2}^{*},  \theta_{1}^{*}\lambda_{0}^{*}).
$$
Hence, $Z_\Theta Z_\Lambda=Z_\Gamma^\dagger$ where $\Gamma$ has the components $(\gamma_0,\gamma_1,\gamma_2)$ just given.  From $Z_{\Gamma}=D_{\Gamma}^{*}H^{\dagger}ZHD_{\Gamma}$, we obtain ${Z_{\Theta}Z_{\Lambda}=Z_{\Gamma}^{\dagger}=D_{\Gamma}^{*}H^{\dagger}Z^{\dagger}HD_{\Gamma}}$. The product observable $Z_\Theta Z_\Lambda$ can consequently also be implemented by a tritter and a detector, but with the detector performing a measurement corresponding to the observable $Z^{\dagger}$ instead of $Z$.

  Violations of Inequality~(\ref{HBI}) by quantum states have been computed in~\cite{Arnault2012},
using observables (among them, product observables) obtained from the ones of
Figure~\ref{pic:hresult1}.  More precisely, the local realistic elements $A_1^2$, $A_1A_2$ and
$A_2^2$ have to be replaced by the three observables
$$
  Z_{\Theta_A}^*, \quad Z^\dagger_{\Gamma_A},\quad  Z_{\Lambda_A}^*
$$
where the $Z_{\Gamma_A}^\dagger$ is a product observable as just described (and $\Theta_A$, $\Lambda_A$ are the parameters corresponding to the configuration of Figure~\ref{pic:hresult1}).  Similarily,
the party Bob has to use three  observables $Z_{\Theta_B}^*$, $Z^\dagger_{\Gamma_B}$, $Z_{\Lambda_B}^*$.

  As mentionned in~\cite{Arnault2012}, the violation factor obtained with the bases obtained from 
Figure~\ref{pic:hresult1} and the GHZ state is $v\simeq1.693$.  Hence, this amount of violation can be
observed using six tritters with detectors.

\subsection{The h3DEB procedure}

As for 3DEB, we denote $A_i$ with $i=0,1,2,3$ the observable parameterized by phase shift triple $(1,\zeta^i,\zeta^{2i})$, and $B_j$ with $j=0,1,2,3$ the observable parameterized by $(1,\zeta^{-j},\zeta^{-2j})$.
\medskip

For each pair $ij$ in the set $\mathcal{C}=\{00,02,22,11,13,33\}$, we note now $A_{ij}$ the product observable $A_{i}A_{j}$ (which is expected to be implemented with a single tritter). 

\begin{enumerate}
\item Alice and Bob obtain an entangled pair of states in the GHZ state.
\item Alice draws randomly a value of $ij$ in $\mathcal{C}$ and performs her measurement in the basis associated to the observable $A_{ij}$ whereas Bob draws randomly a value of $kl$ in $\mathcal{C}$ and performs his measurement in the basis associated to $B_{kl}$.
\item When the pairs $ij$ and $kl$ are equal, the triplets of phase shifts $(\theta_{0}, \theta_{1}, \theta_{2})$ and $(\theta_{0}^{*}, \theta_{1}^{*}, \theta_{2}^{*})$ corresponding to the observables $A_{ij}$ and $B_{kl}$ are related each other by complex conjugation.   Equation~(\ref{CollectTrits}), which has yet been used in the special case where $\theta_j=\theta^j$ for some $\theta$, remains true in the present slightly more general case.  Thus Alice and Bob obtain a new trit for the shared key.
\item For some choices of pairs $ij$ and $kl$ (see the table below) Alice and Bob collect the issues
of their measurements, in order to detect eavesdropping.  Indeed, these pairs correspond to two
configurations of maximal violation of the homogeneous Bell inequality hCHSH-3 given by
Equation~(\ref{HBI}).
\end{enumerate}

These different cases can be summarized in the table :

\begin{table}[h]
\begin{minipage}[h]{0.57\linewidth}
\centering
\begin{tabular}{c | c c c c c c}
\hline\hline
    & $B_{00}$ & $B_{02}$ & $B_{22}$ & $B_{11}$ & $B_{13}$ & $B_{33}$\\
\hline
$A_{00}$ & $k$ &   &    &  $c_{1}$  & $c_{1}$ & $c_{1}$  \\ 
$A_{02}$ &  &  $k$ &    & $c_{1}$   & $c_{1}$ & $c_{1}$  \\
$A_{22}$ &  &   &   $k$ &   $c_{1}$ & $c_{1}$ & $c_{1}$  \\
$A_{11}$ & $c_{2}$ &  $c_{2}$ &  $c_{2}$ &   $k$ &  &   \\
$A_{13}$ & $c_{2}$ &  $c_{2}$ &  $c_{2}$  &    & $k$ &   \\
$A_{33}$ & $c_{2}$ &  $c_{2}$ &   $c_{2}$ &    &  & $k$  \\
\end{tabular}
\label{tab:hresult2}
\caption{Cases for the h3DEB protocol} 
\end{minipage}
\hfill
\begin{minipage}[h]{0.4\linewidth}
$k$ : Alice and Bob obtain a key trit.\par
$c_{i}$ : Alice and Bob collect values for two sets of data which will be used to check hCHSH-3 violation.
\end{minipage}
\end{table}
\vskip-1cm\hbox{}

\subsection{Resistance to noise}

Without the presence of noise, the violation of the inequality hCHSH-3 observed with this protocol
is~$v\simeq1.693$.  By the same argument as the one used for the resistance of 3DEB, we obtain
that our protocol is resistant to noise up to a threshold $F=1-\frac{1}{v} \simeq 0.409$.  This
is better than the resistance of the 3DEB protocol, even when the latter uses the non maximally
entangled state~(\ref{NMEstate}).

\section{Conclusion}
 
Our goal was to improve the noise resistance of the qutrits key distribution protocols.  By using the homogeneous Bell inequality hCHSH-3 which reaches a violation factor $v\simeq1.693$ with the GHZ state, better than for CHSH-3, we obtain a threshold of noise resistance $F\simeq0.409$,  better than the threshold $\simeq0.304$ obtained for 3DEB using the GHZ state~\cite{UptoN16, ViolCHSH3GHZ} and even to the threshold ${\simeq0.341}$ resulting of the use of the non maximally entangled state~(\ref{NMEstate}).

As our inequality hCHSH-3 involves products of observables which become incompatibles for quantum
states, an important fact is the
possibility to implement with slightly modified tritters the single observable corresponding to
these products.  This can be done by replacing the final measurement with observable
$Z=\diag(1,\omega,\omega^2)$ by a measurement with observable
$Z^\dagger=\diag(1,\omega^2,\omega)$.  Physically, this replacement corresponds just to a permutation of the detectors. 

  The gain in noise resistance of our protocol over 3DEB is due to the use of the inequality hCHSH3.
This inequality detects violations of local realism when some measurements are multiplicatively
related.  By using tritters measurements which respect this multiplicative constraints, the parties
running the protocol are able to exploit its larger violation capabilities.

\bigskip
\noindent{\bf Acknowledgement}.  One of the author (Z.A.) was partially supported by Thales Alenia
Space during this work.
\newpage

\bibliographystyle{unsrt}
\bibliography{biblio}

\newpage

\section*{Appendix A : Obtaining the Inequality~(\ref{InegS}) }

\begin{small}
$S=\text{Re}\big((E(A_{1}B_{1})+E(A_{1}B_{2})-E(A_{2}B_{1})+E(A_{2}B_{2})\big)\\
\phantom{espaceplusgrand}+\frac{1}{\sqrt{3}}\text{Im}\big(E(A_{1}B_{1})-E(A_{1}B_{2})-E(A_{2}B_{1})+E(A_{2}B_{2})\big)$
\end{small}

This inequality can be rewritten :

\begin{center}
$S=\text{Re}(U+V)+\frac{1}{\sqrt{3}}\text{Im}(U-V)$
\end{center}

with $U=E(A_{1}B_{1})-E(A_{2}B_{1})+E(A_{2}B_{2})$ and $V=E(A_{1}B_{2})$

For $i, j=0..3$, we have $A_{i}^{2}=A_{i}^{*}$ and $B_{j}^{2}=B_{j}^{*}$, which gives us :

${T=3((\omega^{2}-1)E(A_{1}^{2}B_{1}^{2}) + (\omega-1)E(A_{1}^{2}B_{2}^{2}) + (1-\omega^{2})E(A_{2}^{2}B_{1}^{2}) + (\omega^{2}-1)E(A_{2}^{2}B_{2}^{2}))}$\\
$\phantom{Te}=3((\omega-1)E(A_{1}B_{1}) + (\omega^{2}-1)E(A_{1}B_{2}) + (1-\omega)E(A_{2}B_{1}) + (\omega-1)E(A_{2}B_{2}))$\\
$\phantom{Te}=3((\omega-1)U + (\omega^{2}-1)V)$

$\text{Re}(T)=3\text{Re}((\omega-1)U + (\omega^{2}-1)V)$\\
$\phantom{espacee}=3\text{Re}(\omega U + \omega^{2} V-U-V)$\\
$\phantom{espacee}=3(\text{Re}(\omega U) + \text{Re}(\omega^{2} V)-\text{Re}(U)-\text{Re}(V))$\\
$\phantom{espacee}=3(\text{Re}(\omega) \text{Re}(U)-\text{Im}(\omega) \text{Im}(U)+ \text{Re}(\omega^{2})\text{Re}(V)-\text{Im}(\omega^{2})\text{Im}(V)$\\
$\phantom{espaceeee}-\text{Re}(U)-\text{Re}(V))$

But we also have $\text{Re}(\omega)=\text{Re}(\omega^{2})=-\frac{1}{2}$ and $\text{Im}(\omega)=-\text{Im}(\omega^{2})=\frac{\sqrt{3}}{2}$, which gives us :

$\text{Re}(T)=3(-\frac{3}{2}\text{Re}(U+V) - \frac{\sqrt{3}}{2}\text{Im}(U-V))$\\
$\phantom{espacee}=-\frac{9}{2}(\text{Re}(U+V) + \frac{1}{\sqrt{3}}\text{Im}(U-V))$\\
$\phantom{espacee}=-\frac{9}{2}S$

$S=-\frac{2}{9}\text{Re}(T)$

\end{document}